\documentclass[11pt,draftcls,peerreview,onecolumn]{IEEEtran}

\usepackage[dvips]{graphics}
\usepackage{graphicx}
\usepackage{times}
\usepackage{amsmath}
\usepackage{amssymb}
\usepackage{color}
\usepackage{fancybox}
\newtheorem{lemma}{\bf Lemma}
\newtheorem{theorem}{\bf Theorem}

\newcommand{\bee}{\begin{eqnarray}}
\newcommand{\eee}{\end{eqnarray}}
\newcommand{\be}{\begin{equation}}
\newcommand{\ee}{\end{equation}}

\newcommand{\al}[1]{\begin{align} #1 \end{align}}

\newcommand{\equ}[1]{\begin{equation} #1 \end{equation}}

\newcommand{\mb}{\mathbf}

\newcommand{\nnb}{\nonumber}

\newcommand{\qa}{{\bf a}}

\newcommand{\qh}{{\bf h}}

\newcommand{\qu}{{\bf u}}

\newcommand{\qx}{{\bf x}}
\newcommand{\qy}{{\bf y}}
\newcommand{\qz}{{\bf z}}

\newcommand{\qA}{{\bf A}}

\newcommand{\qE}{{\bf E}}

\newcommand{\qG}{{\bf G}}

\newcommand{\qI}{{\bf I}}

\newcommand{\qQ}{{\bf Q}}

\begin{document}
%
\title{Explicit Solution of Worst-Case Secrecy Rate for MISO Wiretap Channels with Spherical Uncertainty}

\author{\IEEEauthorblockN{Jiangyuan Li and Athina P. Petropulu}\\
\IEEEauthorblockA{Department of Electrical and Computer Engineering\\
Rutgers-The State University of New Jersey, New Brunswick, NJ 08854}
}

\maketitle

\begin{abstract}
\footnote{Work supported by the
National Science Foundation under grant CNS-0905425.}
A multiple-input single-output (MISO) wiretap channel model is considered,
that includes a multi-antenna transmitter, a single-antenna legitimate receiver and a single-antenna eavesdropper.
For the scenario in which spherical uncertainty for both the legitimate and the eavesdropper channels is included,
the problem of finding the optimal input covariance that maximizes the worst-case secrecy rate
subject to a power constraint, is considered, and an explicit expression for the maximum worst-case secrecy rate is provided.
\end{abstract}


\newpage

\section{Introduction}

Wireless physical (PHY) layer based security approaches exploit the physical characteristics of the
wireless channel to enhance the security of communication systems.
The most basic physical layer model that captures
the problem of communication security was proposed in \cite{Wyner}. Later, the Gaussian scalar wiretap channel was studied in \cite{Hellman}.
Recently, the secrecy capacity of multi-antanna wiretap channels has been studied in \cite{Khisti2}, \cite{Khisti3}.
The above work assumes that perfect channel state information (CSI) on the legitimate and the eavesdropper channels.
When ellipsoidal channel uncertainty is included, the worst-case secrecy rate of multiple-input single-output (MISO) cognitive radio network was studied in \cite{Pei-TSP}. In \cite{Pei-TSP}, the worst-case secrecy rate maximization problem is converted to a quasi-convex problem by using the S-procedure to express the channel uncertainty constraint in a linear matrix inequality (LMI) form.
In \cite{Ma}, a MISO channel in the presence of multiple eavesdroppers, each equipped multiple antennas, was considered. Perfect CSI as well as channel uncertainty were considered in \cite{Ma}. As in \cite{Pei-TSP},
 the problem was converted to a semidefinite programming (SDP) problem by using the S-procedure to express the channel uncertainty constraint in LMI form.

In this paper, we consider the same MISO wiretap scenario  as in \cite{Pei-TSP}, \cite{Ma}, except that we consider a special case of the channel uncertainty considered in \cite{Pei-TSP}, \cite{Ma}, i.e., spherical uncertainly. For this case, we derive
an explicit expression of optimal input covariance that maximizes worst-case secrecy rate subject to a power constraint.
In particular, the solution is obtained via finding the eigenvalues of a known $6$-by-$6$ matrix.
The advantage of such an explicit solution lies in:
\begin{itemize}
  \item [1)] Independent of the number of antennas the problem leads to a $6$-by-$6$ matrix whose entries are obtained by direct and simple calculation  from the estimated channel values (see Eq. (\ref{abcdr}) and (\ref{G})).
  The computation time for finding the eigenvalues of such $6$-by-$6$ matrix is not affected by the size of problem.
       In contrast, for the existing methods (\cite{Pei-TSP}, \cite{Ma}), the computation time for the iteration algorithm is affected significantly by the size of problem.
  \item [2)] The structure of the optimal input covariance matrix is explicitly given (see Eq. (\ref{opti-u})).
  \item [3)] A (simple) necessary and sufficient condition for a positive worst-case secrecy rate is explicitly given (see Theorem \ref{Theo:1}).
\end{itemize}

{\em Notation}- Upper case and lower case bold symbols denote matrices and vectors, respectively.
Superscripts $\ast$, $T$ and $\dagger$
denote respectively conjugate, transposition and conjugate transposition.
${\mb A}\succeq 0$ means that the matrix $\qA$ is Hermitian positive semi-definite.
$|a|$ denotes the absolute value of $a$, while $\|\qa\|$ denotes Euclidean norm of the vector $\qa$.
$\mathrm{eig}(\qA)$ denotes the eigenvalues of the matrix $\qA$.

\section{System Model and Problem Statement}\label{sec:2}

Consider a Gaussian MISO wiretap channel, which includes a transmitter is equipped with $n_T$ antennas, and
a legitimate receiver and an eavesdropper, each equipped with a single antenna.
The received signals at the legitimate receiver and the eavesdropper are respectively given by
\equ{
y_R = \qh_R^\dagger\qx+v_R, \ \mathrm{and}\ y_E = \qh_E^\dagger\qx+v_E\label{system}
}
where $\qx$ is the $n_T\times 1$ transmitted signal vector with zero mean and covariance matrix $\qQ=\mathbb{E}\{\qx\qx^\dagger\}$;
$\qh_R$, $\qh_E$ are respectively channel vectors between the transmitter and legitimate receiver,
and between the transmitter and eavesdropper;
$v_R$, $v_E$ are the noises at the legitimate receiver and the eavesdropper with zero means and unit variances, respectively.

We consider the scenario in which spherical channel uncertainty, i.e., \cite[\S18.2.4]{Bengtsson}
\al{
&\qh_R \in \Omega_R = \{\qh_R\big\vert\, \|\qh_R - \bar{\qh}_R\| \le \epsilon_R\},\\
&\qh_E \in \Omega_E = \{\qh_E\big\vert\, \|\qh_E - \bar{\qh}_E\| \le \epsilon_E\}
}
where $\bar{\qh}_R$ and $\bar{\qh}_E$ are the estimated values, $\epsilon_R$ and $\epsilon_E$ are the estimated error bounds.
Let $P$ be the transmitted power budget.
We investigate the maximum worst-case secrecy rate defined as \cite{Pei-TSP}
\equ{
R_s = \max_{\qQ\succeq 0,\, \mathrm{Tr}(\qQ) \le P}\ \min_{\qh_R \in \Omega_R, \, \qh_E \in \Omega_E}\ R_s(\qQ, \qh_R, \qh_E) \label{worstcase}
}
where $R_s(\qQ, \qh_R, \qh_E) = \log(1+\qh_R^\dagger\qQ\qh_R) - \log(1+\qh_E^\dagger\qQ\qh_E)$.
We aim to find the optimal input covariance matrix $\qQ^\star$.
A necessary condition to ensure $R_s>0$ is $0 \notin \Omega_R$, i.e.,
\equ{
\|\bar{\qh}_R\| > \epsilon_R.\label{CondPosiCs}
}
We assume that the condition (\ref{CondPosiCs}) holds in this paper.

Since $\log(\cdot)$ is an increasing function, (\ref{worstcase}) is equivalent to
\equ{
\max_{\qQ\succeq 0,\, \mathrm{Tr}(\qQ) \le P}\ \min_{\qh_R \in \Omega_R, \, \qh_E \in \Omega_E}\ \frac{1+\qh_R^\dagger\qQ\qh_R}{1+\qh_E^\dagger\qQ\qh_E}.\label{worstcase1}
}
Let $\tau^\star$ be its optimal objective value. Certainly, $\tau^\star > 1$ is equivalent to $R_s>0$.

\section{Explicit Solution of Optimal Input Covariance Matrix}\label{OptInpCov}

One can show that if $\tau^\star>1$ ($R_s>0$), then there exists a rank one matrix, to be denoted by $\qQ^\star$.
The proof can be found in \cite{Ma}. With this, we assume $\tau^\star > 1$ and find the rank one $\qQ^\star$.
If such rank one $\qQ^\star$ achieves $R_s>0$, $\qQ^\star$ is indeed the solution of the problem. Otherwise, $R_s=0$.

Since $\tau^\star>1$,  it is easy to verify that $\mathrm{Tr}(\qQ^\star)=P$.
Then, we let $\qQ=P\qu\qu^\dagger$ and rewrite (\ref{worstcase1}) as
\equ{
\max_{\|\qu\|^2 = 1}\ \frac{1+P\min\limits_{\qh_R \in \Omega_R} \, |\qu^\dagger\qh_R|^2}
{1+P\max\limits_{\qh_E \in \Omega_E} \, |\qu^\dagger\qh_E|^2}.\label{worstcase1-RankOne}
}

Let us denote
\al{
&a=P\|\bar{\qh}_E\|^2, \ \ b=P\|\bar{\qh}_R\|^2, \ \ c=\frac{\epsilon_R}{\|\bar{\qh}_R\|}, \ \ d=\frac{\epsilon_E}{\|\bar{\qh}_E\|}, \nnb\\
&r=\frac{|\bar{\qh}_E^\dagger\bar{\qh}_R|}{\|\bar{\qh}_R\|\, \|\bar{\qh}_E\|},
\ \ \mathrm{and}\ \ z_0=\max(c, \sqrt{1-r^2}\, ). \label{abcdr}
}
Let $z^\star$ be the solution of
\equ{
\max_{z_0 \le z \le 1} \bigg[g(z) = \frac{1 + b(z - c)^2}
{1 + a\big(rz - \sqrt{1-r^2}\sqrt{1-z^2} + d\big)^2}\bigg]. \label{Opt7}
}
The main result of this paper is given in the following theorem.

\medskip

\begin{theorem}\label{Theo:1} \mbox{}
\begin{itemize}
  \item [i)] {\em The sufficient and necessary condition for $R_s > 0$ is that
\equ{
\! \left\{\!\! \begin{array}{l}
         \sqrt{1-r^2}\, z_0 + \big(r-\sqrt{b/a}\ \big)\sqrt{1-z_0^2} \, \ge 0 \\
         \sqrt{1 \! - \! r^2}\sqrt{1 \! - \! z_0^2} - \big(r \! - \!\! \sqrt{b/a}\ \big)z_0 > c\sqrt{b/a}+d
       \end{array}
\right. \!\!\! \label{cond1}
}
\equ{
\!\!\! \mathrm{or}\ \left\{\begin{array}{l}
         \sqrt{1-r^2}\, z_0 + \big(r-\sqrt{b/a}\ \big)\sqrt{1-z_0^2} \, < 0 \\
         \sqrt{1-r^2 + (\sqrt{b/a} - r)^2} \,  > c\sqrt{b/a}+d
       \end{array}
\right. . \label{cond2}
}
}
  \item [ii)] {\em Assume that (\ref{cond1}) or (\ref{cond2}) holds. In other words, $\tau^\star>1$. The solution of (\ref{worstcase1-RankOne}) is
\equ{
\qu^\star \! = \! \sqrt{\frac{1-{z^\star}^2}{1-r^2}} \frac{\bar{\qh}_E}{\|\bar{\qh}_E\|} - \Big(\sqrt{\frac{1-{z^\star}^2}{1-r^2}}\, + \frac{z^\star}{r}\Big)
\frac{(\bar{\qh}_R^\dagger\bar{\qh}_E)\bar{\qh}_R}{\|\bar{\qh}_R\|^2\|\bar{\qh}_E\|}. \label{opti-u}
}
It can be seen from (\ref{opti-u}) that $\qu^\star$ is a linear combination of $\bar{\qh}_R$ and $\bar{\qh}_E$.
The maximum worst-case secrecy rate is
\equ{
R_s = \log \frac{1 + b(z^\star - c)^2}
{1 + a\big(rz^\star - \sqrt{1-r^2}\sqrt{1-{z^\star}^2} + d\big)^2}. \label{Rs-z}
}
}
\end{itemize}
\end{theorem}
{\em Remarks:} Let us gain some insight into the conditions for $R_s>0$. Obviously, (\ref{cond1}) and (\ref{cond2}) are both independent of $P$.
In other words, if neither of these two conditions holds,
$R_s>0$ cannot be achieved even when infinite power is used. Roughly speaking, this may occur if the estimation is not good enough
($c$, $d$ is relatively large),
or the legitimate channel is worse than the eavesdropper channel, i.e., $\|\bar{\qh}_R\| < \|\bar{\qh}_E\|$.

\medskip

{\bf Proof of Theorem \ref{Theo:1}}

\medskip

For the optimal $\qu$, consider the two subproblems in (\ref{worstcase1-RankOne}):
\al{
&\min_{\qh_R \in \Omega_R}\, |\qu^\dagger\qh_R|^2, \label{Opt2innerA} \\
\mathrm{and} \ &\max_{\qh_E \in \Omega_E}\, |\qu^\dagger\qh_E|^2.\label{Opt2innerB}
}
Their optimal objective values can be obtained in closed form, given by
$(|\qu^\dagger\bar{\qh}_R| - \epsilon_R)^2$ and $(|\qu^\dagger\bar{\qh}_E| + \epsilon_E)^2$ respectively.
Moreover, since $\tau^\star > 1$, it holds that $|\qu^\dagger\bar{\qh}_R| > \epsilon_R$.
More details can be found in Appendix \ref{ProofLem:ClosedFormSolution}.
From this result, (\ref{worstcase1-RankOne}) is equivalent to
\equ{
\max_{\|\qu\|^2=1, \ |\qu^\dagger\bar{\qh}_R| > \epsilon_R}\ \frac{1 + P(|\qu^\dagger\bar{\qh}_R| - \epsilon_R)^2}{1 + P(|\qu^\dagger\bar{\qh}_E| + \epsilon_E)^2}\label{Opt4}.
}
We can reduce (\ref{Opt4}) to a problem of one variable.
To this end, let $z=|\qu^\dagger\bar{\qh}_R|/\|\bar{\qh}_R\|$.
The domain of $z$ is
\equ{
\frac{\epsilon_R}{\|\bar{\qh}_R\|} < z \le 1, \label{zdom}
}
which follows from Cauchy's inequality and the constraint $|\qu^\dagger\bar{\qh}_R| > \epsilon_R$.
It can be seen that the condition (\ref{CondPosiCs}) ensures a non-empty domain of $z$.
Obviously, for fixed $z$ (i.e., $|\qu^\dagger\bar{\qh}_R|$ is fixed at $\|\bar{\qh}_R\|z$),
$|\qu^\dagger\bar{\qh}_E|$ should be minimized.
Keeping this in mind, let
\al{
\psi(z) \triangleq &\min_{\qu} \ |\qu^\dagger\bar{\qh}_E|\\
&\mathrm{s.t.}\quad \|\qu\|^2=1, \ \mathrm{and}\ |\qu^\dagger\bar{\qh}_R|=\|\bar{\qh}_R\|z.\nnb
}
Then, we rewrite (\ref{Opt4}) as
\equ{\label{Opt6}
\max_z\ \frac{1 + P(\|\bar{\qh}_R\|z - \epsilon_R)^2 }{1 + P(\psi(z) + \epsilon_E)^2} \qquad
\mathrm{s.t.}\quad \frac{\epsilon_R}{\|\bar{\qh}_R\|} < z \le 1.
}
The function $\psi(z)$ is obtained in closed form using the following lemma.

\medskip

\begin{lemma}\label{Lem:psi_z}
{\em
Let $\qa$ and $\mb{b}$ be (known) unit-norm vectors with $r=|\mb{b}^\dagger\qa| < 1$,
and $0\le q\le 1$. The problem
\al{
&\min_{\qu} \ \qu^\dagger\mb{b}\mb{b}^\dagger\qu\label{lem1op1}\\
&\mathrm{s.t.}\quad \qu^\dagger\qa\qa^\dagger\qu=q, \ \mathrm{and}\ \|\qu\|^2=1\nnb
}
has an optimal objective value
\equ{
\left\{\begin{array}{cc}
         (r\sqrt{q} - \sqrt{1-r^2}\sqrt{1-q}\,)^2 & \mathrm{if}\ q \ge 1-r^2 \\
         0 & \mathrm{if}\ q \le 1-r^2
       \end{array}
\right..
}
Moreover, if $q \ge 1-r^2$, then the optimal $\qu$ is given by
\equ{
\qu^\star = - \bigg(\sqrt{\frac{1-q}{1-r^2}}\,+ \frac{\sqrt{q}}{r}\bigg)(\qa^\dagger\mb{b})\qa + \sqrt{\frac{1-q}{1-r^2}}\, \mb{b}. \label{opt-u}
}
}
\end{lemma}
The proof is given in Appendix \ref{proofLem:psi_z}. The proof is similar to the one of Lemma 2 in \cite{Li-TSP}.
\medskip

We assume that $\bar{\qh}_R$ and $\bar{\qh}_E$ are linearly independent (otherwise, the problem is reduced and simpler).
Let $\qa = \bar{\qh}_R/\|\bar{\qh}_R\|$, $\mb{b}=\bar{\qh}_E/\|\bar{\qh}_E\|$,
$r=|\bar{\qh}_E^\dagger\bar{\qh}_R|/(\|\bar{\qh}_R\| \, \|\bar{\qh}_E\|)$, $q=z^2$.
It follows from Lemma \ref{Lem:psi_z} that
\equ{
\psi(z)  =  \Big\{\begin{array}{cl}
                   \|\bar{\qh}_E\|(rz - \sqrt{1-r^2}\sqrt{1-z^2}\,) & z \ge \sqrt{1-r^2} \\
                   0 & \mathrm{otherwise}
                 \end{array}
\Big. .
}

Note that in (\ref{Opt6}), $(|\bar{\qh}_R|z - \epsilon_R)^2$ is an increasing function of $z$.
On the other hand, $\psi(z')=0$ for any $z' < \sqrt{1-r^2}\,$.
Thus, it holds that $z^\star\ge \sqrt{1-r^2}\, $ (otherwise, there exists $z^\star < z' < \sqrt{1-r^2}\,$ such that
$z'$ achieves a larger objective value. But this contradicts the optimality of $z^\star$).
With this, we rewrite (\ref{Opt6}) as (\ref{Opt7}).
Here we replace the constraint $z > c$ by $z \ge c$ for convenience.
Furthermore, from (\ref{opt-u}), we obtain (\ref{opti-u}).

From the objective in (\ref{Opt7}), the sufficient and necessary condition for $R_s > 0$ is that there exists a $z_0 \le z \le 1$ such that
$\sqrt{b}\, (z - c) > \sqrt{a}\, \big(rz - \sqrt{1-r^2}\sqrt{1-z^2} + d\big)$, or equivalently
\al{
&\max_{z_0 \le z \le 1} \big[g_1(z)=\sqrt{1-r^2}\sqrt{1-z^2} \, - \big(r-\sqrt{b/a}\ \big)z \big] \nnb\\
& > c\sqrt{b/a}+d. \label{condition}
}
Consider the optimization problem in the left hand side of (\ref{condition}).
The derivative $g_1'(z)=-z\sqrt{1-r^2}/\sqrt{1-z^2} \, - \big(r-\sqrt{b/a}\ \big)$ is
a strictly decreasing function of $z$. On the other hand, $g_1'(z) \to -\infty$ as $z \to 1$.
Thus, if $g_1'(z_0)\le 0$, then $z_0$ is the solution; if $g_1'(z_0) > 0$, then the optimal $z$ is the unique point with $z_0<z<1$ and $g_1'(z)=0$.
From this fact, we obtain (\ref{cond1}) or (\ref{cond2}).
This completes the proof.

\subsection{Determining $z^\star$}

Note that the derivative $g'(z)\to -\infty$
as $z \to 1$. Thus $z^\star$ is either $z_0$ or one of the feasible points (if any) with $g'(z)=0$.
We introduce the bijective transform
\equ{
z = \frac{2x}{1+x^2}, \ 0\le x\le 1 \label{transform}
}
to rewrite (\ref{Opt7}) as
\al{
&\max_x\ \Big[F(x)=\frac{p_0x^4+p_1x^3+p_2x^2+p_1x+p_0}{q_0x^4+q_1x^3+q_2x^2+q_3x+q_4}\Big] \label{Opt8}\\
&\mathrm{s.t.}\quad \frac{1- \sqrt{1-z_0^2}}{z_0} \le x \le 1 \nnb
}
where $p_0 = 1+bc^2$, $p_1 = -4bc$, $p_2 = 4b+2bc^2+2$,
$q_0 = 1+a(\sqrt{1-r^2} +d)^2$, $q_1 = 4ar(\sqrt{1-r^2} +d)$,
$q_2 = 2-2a+6ar^2+2ad^2$, $q_3 = 4ar(-\sqrt{1-r^2}+d)$,
$q_4 = 1+a(-\sqrt{1-r^2}+d)^2$.
The solution $x^\star$ should be $(1- \sqrt{1-z_0^2}\, )/z_0$ or one of the feasible points (if any)
with $F'(x)=0$. One can rewrite $F'(x)=0$ as an equation of six degree
\equ{
a_0x^6 + a_1x^5 + a_2x^4 + a_3x^3 + a_4x^2 + a_5x + a_6 = 0 \label{equ-six}
}
where $a_0 = p_1q_0-p_0q_1$, $a_1 = 2p_2q_0-2p_0q_2$,
$a_2 = 3p_1q_0+p_2q_1-p_1q_2-3p_0q_3$,
$a_3 = 4p_0q_0+2p_1q_1-2p_1q_3-4q_4p_0$,
$a_4 = 3p_0q_1+p_1q_2-p_2q_3-3q_4p_1$,
$a_5 = 2p_0q_2-2q_4p_2$,
$a_6 = p_0q_3-q_4p_1$.
All the six roots of the equation (\ref{equ-six}) equal the eigenvalues $\lambda_i$ of the following $6\times 6$ matrix \cite[Ch. 6]{McNamee}
\equ{
\qG = \left(
        \begin{array}{llllll}
          -\frac{a_1}{a_0} & -\frac{a_2}{a_0} & -\frac{a_3}{a_0} & -\frac{a_4}{a_0} & -\frac{a_5}{a_0} & -\frac{a_6}{a_0} \\
           &  &  \qI_5 &  &  & 0_{5\times 1} \\
        \end{array}
      \right). \label{G}
}
Thus, the optimal $x$ satisfies (if there exists several feasible candidates for $x^\star$, the best one is chosen)
\equ{
x^\star \in \Big\{\frac{1- \sqrt{1-z_0^2}}{z_0}, \quad \mathrm{eig}(\qG)\Big\}.
}
Once $x^\star$ is obtained, we have
\equ{
z^\star = \frac{2x^\star}{1+{x^\star}^2}.
}

\section{Numerical Simulations}\label{sec:sim}

In this section we provide some examples to illustrate the result.
We focus on how to obtain the explicit solution. For more numerical simulations
on different estimated channel values, e.g., $(\epsilon_R, \epsilon_E)$ please refer to \cite{Pei-TSP} and \cite{Ma}.

First consider a MISO wiretap channel with $n_T=4$ antennas. We set $(\epsilon_R, \epsilon_E)=(10^{-2}, 10^{-2})$, $P= 5\, \mathrm{dB}$ and
\equ{
\bar{\qh}_R \! = \!\! {\small
                \left(\!\!\!
                \begin{array}{r}
                   -1.0301 + 0.3060\mathrm{i} \\
                   -0.0162 + 0.5618\mathrm{i} \\
                   0.7134 - 0.1504\mathrm{i} \\
                   1.0488 + 0.1086\mathrm{i} \\
                \end{array}\!\!
              \right)
              }, \
\bar{\qh}_E \! = \!\! {\small
                \left(\!\!\!
                \begin{array}{r}
                   -0.3475 - 0.0816\mathrm{i} \\
                  0.3662 - 0.1442\mathrm{i} \\
                  0.2450 - 0.4282\mathrm{i} \\
                  0.2369 + 0.2346\mathrm{i} \\
                \end{array}\!\!
              \right)
              }. \nnb
}
According to (\ref{abcdr}), we obtain that $a=2.01390,    b=9.84720, c=0.0566687, d=0.0125309,    r=0.540848, z_0=0.841120$.
Then, according to Theorem \ref{Theo:1} i), we calculate
$\sqrt{1-r^2}\, z_0 + \big(r-\sqrt{b/a}\ \big)\sqrt{1-z_0^2} \, = -0.1959 < 0$,
$\sqrt{1-r^2 + (\sqrt{b/a} - r)^2} \,  - ( c\sqrt{b/a}+d)  = 1.8452 > 0$.
Thus, (\ref{cond2}) holds and hence $R_s > 0$.

Then, we have
\equ{
\mathrm{eig}(\qG) \! = \! {\small \big[47.92, -1.46, 0.0016 \pm 0.99\mathrm{i},
0.6741, -0.016 \big] }. \nnb
}
There is only one feasible $x=0.6741$ among $\mathrm{eig}(\qG)$.
Then, we obtain $x^\star=0.6741$ and hence $z^\star = 0.9270$.

\begin{figure}[hbtp]
\centering
\includegraphics[width=4in]{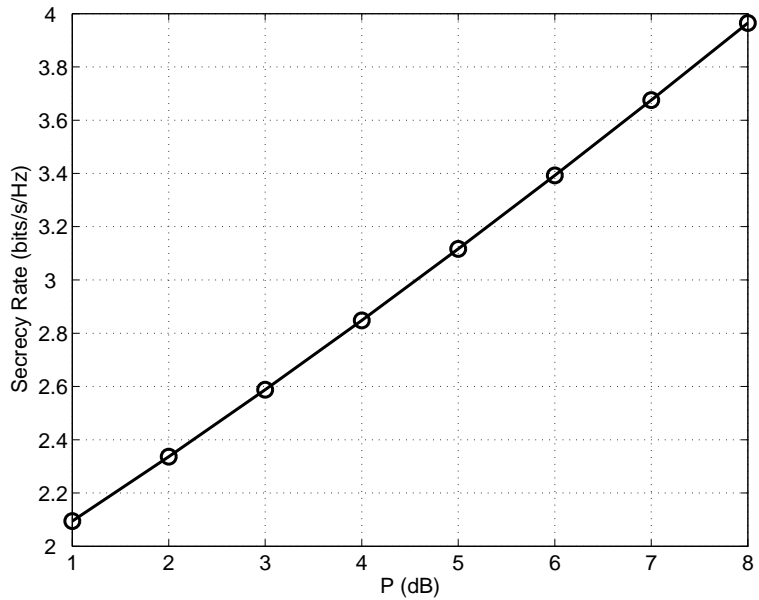}
\caption{Worst-case secrecy rate versus power, $P$, of a MISO wiretap channel with $4$
transmit antennas. $(\epsilon_R, \epsilon_E)=(10^{-2}, 10^{-2})$.}
\label{fig:RsSNR}
\end{figure}

Finally, according to (\ref{opti-u}) and (\ref{Rs-z}), we have $\qu^\star = [0.4692 - 0.3024\mathrm{i}, \   0.1854 - 0.4521\mathrm{i}, \ -0.3258 - 0.1020\mathrm{i}, \  -0.5655 + 0.1153\mathrm{i}]^T$ and
\equ{
R_s = 3.1162 \ \mathrm{(bits/s/Hz)}. \nnb
}
Fig. \ref{fig:RsSNR} plots the secrecy rate for different power $P$.

\smallskip

Second, we give an example in which $R_s>0$ cannot be achieved.
We set $(\epsilon_R, \epsilon_E)=(0.05, 0.05)$, and
$\bar{\qh}_R=[0.1216 \! + \! 0.0118\mathrm{i}, 0.0106 \! - \! 0.0316\mathrm{i},  -0.0856 \! - \! 0.1063\mathrm{i},
0.2241 \! - \! 0.0216\mathrm{i}]$,
$\bar{\qh}_E = [0.3599 \! + \! 0.0174\mathrm{i},   0.1655 \! - \! 0.1923\mathrm{i},  -0.2323 \! - \! 0.4065\mathrm{i}, 0.7313 \! - \! 0.2272\mathrm{i}]$. Neither of (\ref{cond1}) and (\ref{cond2}) holds. Thus, $R_s>0$ cannot be achieved even when infinite power is used.

\section{Conclusion}\label{Sec:Conc}

We study the problem of finding the optimal input covariance that maximizes the worst-case secrecy rate of a MISO wiretap channel under channel uncertainty, subject to a power constraint.
We show that the optimal input covariance can be obtained via finding the eigenvalues of a known $6$-by-$6$ matrix.

\appendices

\section{Closed Form Solutions of (\ref{Opt2innerA}) and (\ref{Opt2innerB})}\label{ProofLem:ClosedFormSolution}

\underline{First}, we solve (\ref{Opt2innerB}). Let $\qh_E = \bar{\qh}_E + \epsilon_E\qy$.
The constraint $\qh_E\in \Omega_E$ becomes $\|\qy\|\le 1$. Thus, (\ref{Opt2innerB}) is equivalent to
\equ{
\max_{\|\qy\|\le 1}\, |\qu^\dagger\bar{\qh}_E + \epsilon_E\qu^\dagger\qy|^2.\label{Opt2innerBApp-1}
}
First, $|\qu^\dagger\bar{\qh}_E| + \epsilon_E$ is an upper bound of $|\qu^\dagger\bar{\qh}_E + \epsilon_E\qu^\dagger\qy|$
which can be easily verified from the triangle inequality and Cauchy's inequality.
Moreover, this upper bound can be achieved: $\qy   = (\qu^\dagger\bar{\qh}_E/|\qu^\dagger\bar{\qh}_E|)\qu$
(if $\qu^\dagger\bar{\qh}_E=0$, then $\qy=\qu$).
Thus, the optimal objective value of (\ref{Opt2innerBApp-1})
is $(|\qu^\dagger\bar{\qh}_E| + \epsilon_E)^2$.

\underline{Second}, we solve (\ref{Opt2innerA}). Let $\qh_R=\bar{\qh}_R+\epsilon_R\qx$.
The constraint $\qh_R\in\Omega_R$ becomes $\|\qx\|\le 1$. Thus, (\ref{Opt2innerA}) is equivalent to
\equ{
\min_{\|\qx\|\le 1}\, |\qu^\dagger\bar{\qh}_R + \epsilon_R\qu^\dagger\qx|^2. \label{Opt2innerAApp-1}
}
Since $\tau^\star>1$, it holds that the optimal objective value of (\ref{Opt2innerAApp-1}) is greater than zero.
Note that the objective value of (\ref{Opt2innerAApp-1}) is zero at the point $\qx_1 = -(\qu^\dagger\bar{\qh}_R/\epsilon_R)\qu$.
Thus, $\qx_1$ must be infeasible and hence $|\qu^\dagger\bar{\qh}_R|> \epsilon_R$.
With this, we can show that $|\qu^\dagger\bar{\qh}_R| - \epsilon_R$ is a lower bound of
$|\qu^\dagger\bar{\qh}_R + \epsilon_R\qu^\dagger\qx|$ which can be easily verified
from the reverse triangle inequality and Cauchy's inequality.
Moreover, this lower bound can be achieved: $\qx=-(\qu^\dagger\bar{\qh}_R/|\qu^\dagger\bar{\qh}_R|)\qu$.
Thus, the optimal objective value of (\ref{Opt2innerAApp-1}) is $(|\qu^\dagger\bar{\qh}_R| - \epsilon_R)^2$.
This completes the proof.

\section{Proof of Lemma \ref{Lem:psi_z}}\label{proofLem:psi_z}

First, we show that the optimal objective value is zero if and only if $q\le 1-r^2$.
Denote the null space of $\mb{b}$ as $\qE_b$ with
$\qE_b^\dagger\qE_b=\qI$ and $\qE_b\qE_b^\dagger=\qI-\mb{b}\mb{b}^\dagger$.
Then $\mb{b}^\dagger\qu=0$ if and only if there exists a vector $\qz$ such that $\qu = \qE_b\qz$
which, when inserted into $\qu^\dagger\qa\qa^\dagger\qu=q$ and $\|\qu\|^2=1$, results in that $\|\qz\|^2=1$ and $\qz^\dagger\qE_b^\dagger\qa\qa^\dagger\qE_b\qz=q$.
Note that $0\le \qz^\dagger\qE_b^\dagger\qa\qa^\dagger\qE_b\qz \le \lambda_{\max}(\qE_b^\dagger\qa\qa^\dagger\qE_b)=\qa^\dagger\qE_b\qE_b^\dagger\qa=\qa^\dagger(\qI-\mb{b}\mb{b}^\dagger)\qa=1-r^2$.
In other words, $\|\qz\|^2=1$ and $\qz^\dagger\qE_b^\dagger\qa\qa^\dagger\qE_b\qz=q$ hold
if and only if $q\le 1-r^2$. The desired result follows.

Second, we consider the case $q>1-r^2$.
The optimal $\qu$ is a linear combination of $\qa$ and $\mb{b}$, which follows from its optimality condition $\mb{b}\mb{b}^\dagger\qu-\mu_1\qa\qa^\dagger\qu-\mu_2\qu=0$ or equivalently $\mu_2\qu=(\mb{b}^\dagger\qu)\mb{b}-(\mu_1\qa^\dagger\qu)\qa$ where $\mu_1$ and $\mu_2$ are multipliers (Obviously, $\mu_2\ne 0$ since $r<1$). From this result, we let $\qu=c_1\qa+c_2\mb{b}$. Since $e^{\mathrm{i}\omega}\qu$ (for any real $\omega$) satisfies the constraints and attains the same objective value as $\qu$, we can restrict $c_2\ge 0$. Inserting $\qu=c_1\qa+c_2\mb{b}$ into the constraints and objective, results in
\al{
|c_1|^2+c_2^2+c_1^\ast c_2\qa^\dagger\mb{b}+c_1c_2\mb{b}^\dagger\qa&=1,\label{c1c2eq1}\\
|c_1|^2+c_2^2r^2+c_1^\ast c_2\qa^\dagger\mb{b}+c_1c_2\mb{b}^\dagger\qa&=q,\label{c1c2eq2}\\
\mathrm{and}\ \qu^\dagger\mb{b}\mb{b}^\dagger\qu=1-|c_1|^2(1-r^2).\label{obj}
}
From (\ref{obj}), we need to maximize $|c_1|^2$. From (\ref{c1c2eq1}) and (\ref{c1c2eq2}), we get $c_2^2(1-r^2)=1-q$ which leads to $c_2=\sqrt{(1-q)/(1-r^2)}\, $. By denoting $c_1=|c_1|e^{\mathrm{i}\theta}$ where $\theta$ is the argument of $c_1$, we can rewrite (\ref{c1c2eq1}) as
\equ{
|c_1|^2+|c_1|c_2r (e^{-\mathrm{i}(\phi+\theta)} +e^{\mathrm{i}(\phi+\theta)})+(c_2^2-1)=0\label{c1c2eq1-b}
}
where $\phi$ is the argument of $\mb{b}^\dagger\qa$.
This is a quadratic equation of one variable $|c_1|$.
It is not difficult to show that the optimal $\theta = \pi -\phi$, and the optimal $|c_1|=c_2r+\sqrt{q}$.
Thus, the optimal $c_1=(c_2r+\sqrt{q}\, )e^{\mathrm{i}(\pi-\phi)}$, and hence
\equ{
\qu = - \bigg(\sqrt{\frac{1-q}{1-r^2}}\,+ \frac{\sqrt{q}}{r}\bigg)(\qa^\dagger\mb{b})\qa + \sqrt{\frac{1-q}{1-r^2}}\, \mb{b}.
}
Here we have used the fact that, $e^{\mathrm{i}(\pi-\phi)}=-e^{-\mathrm{i}\phi}=-\qa^\dagger\mb{b}/r$.
Further, from (\ref{obj}), we obtain
\al{
\qu^\dagger\mb{b}\mb{b}^\dagger\qu &=1-(c_2r+\sqrt{q})^2(1-r^2)\nnb\\
&=(r\sqrt{q} - \sqrt{1-r^2}\sqrt{1-q}\,)^2. \label{AppObj}
}
This completes the proof.

\end{document}